# DAΦNE DEVELOPMENTS FOR THE KLOE-2 EXPERIMENTAL RUN

C. Milardi for the DAΦNE Commissioning Team[*], INFN-LNF, Frascati, Italy.

*Abstract*

Recently the peak luminosity achieved on the DAΦNE collider has been improved by almost a factor three by implementing a novel collision scheme based on large Piwinski angle and Crab-Waist. This encouraging result opened new perspectives for physics research and a new run with the KLOE-2 detector has been scheduled to start by spring 2010. The KLOE-2 installation is a complex operation requiring a careful design effort and a several months long shutdown. The high luminosity interaction region has been deeply revised in order to take into account the effect on the beam caused by the solenoidal field of the experimental detector and to ensure background rejection. The shutdown has been also used to implement several other modifications aimed at improving beam dynamics: the wiggler poles have been displaced from the magnet axis in order to cancel high order terms in the field, the feedback systems have been equipped with stronger power supplies and more efficient kickers and electrodes have been inserted inside the wiggler and the dipole vacuum chambers, in the positron ring, to avoid the e-cloud formation. A low level RF feedback has been added to the cavity control in both rings.

## INTRODUCTION

DAΦNE [1] the Frascati electron positron collider began steady operations in 2001 and in the next seven years provided high K meson rates, at the energy of the Φ resonance, to three different experiments, which logged ~ 4.4 fb$^{-1}$ total integrated luminosity in dedicated runs. The KLOE detector collected the largest part of these data, 3.0 fb$^{-1}$. In the same period the collider performances have been significantly improved by several progressive upgrades and a wide program of machine measurements and studies has been undertaken aimed at pointing out the factors limiting the maximum achievable luminosity. This activity largely contributed to define a proposal for an inventive collision scheme based on large Piwinski angle and Crab-Waist (CW) compensation of the beam-beam interaction [2].

The novel approach to collision has been implemented on DAΦNE [3] during a six months shutdown already planned to install a compact detector without solenoidal field offering a simplified environment to test the new configuration.

*D. Alesini, M.E. Biagini, C. Biscari, R. Boni, M. Boscolo, F. Bossi, B. Buonomo, A. Clozza, G. Delle Monache, T. Demma, E. Di Pasquale, G. Di Pirro, A. Drago, M. Esposito, A. Gallo, A. Ghigo, S. Guiducci, C. Ligi, F. Marcellini, G. Mazzitelli, C. Milardi, L. Pellegrino, M. Preger, L. Quintieri, P. Raimondi, R. Ricci, U. Rotundo, C. Sanelli, M. Serio, F. Sgamma, B. Spataro, A. Stecchi, A. Stella, S. Tomassini, C. Vaccarezza, M. Zobov, INFN-LNF, Frascati; E. Levichev, S. Nikitin, P. Piminov, D. Shatilov, BINP SB RAS, Novosibirsk; S. Bettoni, CERN, Geneva.

The results obtained [4] are remarkable: the luminosity achieved a $4.53 \cdot 10^{32}$ cm$^{-2}$ s$^{-1}$ peak value with a maximum daily integrated luminosity of ~15 pb$^{-1}$, eventually after 18 months of steady operation ~2.4 fb$^{-1}$ have been delivered to the experiment, see Fig. 1. At the same time the effectiveness of the Crab-Waist collisions has been unquestionably proved by several independent tests and beam measurements [5].

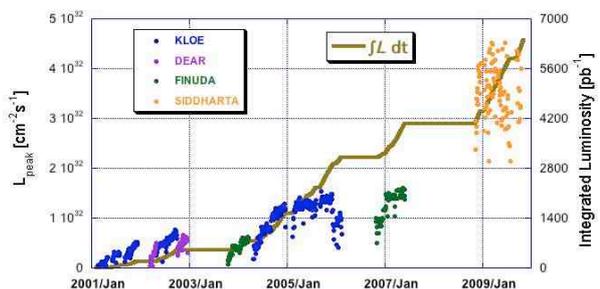

Figure 1: Peak (dots) and integrated (line) luminosity acquired on DAΦNE by the four different experiments.

This relevant jump ahead has been attained despite the collider performances were still affected by several limiting factors. The maximum current storable in the positron beam was 1.2 A due to e-cloud instability. The beam lifetime at maximum current in collision was ~ 550 sec. Machine uptime was affected by hardware reliability due to aged equipment.

## MOTIVATION FOR THE KLOE-2 RUN

The interest in undertaking a new physics run with an upgraded KLOE detector, KLOE-2 [6], is motivated by physical considerations. These require delivering a high statistics data sample to a detector having an improved sensitivity; moreover compatibility of the CW collision scheme with a large detector with a solenoidal field will be tested.

KLOE is a multipurpose experiment devoted, mainly, to study decays of K mesons as well as several hadronic physics and low energy quantum chromodynamics phenomena. The detector consists of a large cylindrical drift chamber, ~3.5 m long and 2 m in radius, surrounded by a lead-scintillating fibre electromagnetic calorimeter. A superconducting coil around the detector provides a magnetic field of 0.52 T. With respect to the original design, KLOE-2 has additional detector layers, including new tracking and calorimeter devices. It also extends its investigation capabilities to the study of gamma-gamma reactions, by means of dedicated detectors [7] tagging the scattered electron and positron, typical of those events.

The maximum daily integrated luminosity measured with Crab-Waist collision is a factor 50% higher than the best achieved during the past KLOE run [8]. This value is consistent with the average daily integrated luminosity per month (~10 pb$^{-1}$) delivered in the final part of the run operating the collider in a moderate injection regime to avoid excessive background in the detector. Tests carried out adopting a continuous injection regime, compatible with the KLOE-2 data-taking, yielded a hourly integrated luminosity L$_{fl\ hour}$ ~1.0 pb$^{-1}$ [9]. Scaling this best integrated luminosity measured over two hours it is reasonable to expect a daily integrated luminosity larger than 20 pb$^{-1}$, and assuming 80% collider uptime a monthly integrated luminosity of ~0.5 fb$^{-1}$.

## THE KLOE-2 INTERACTION REGION

Integrating the high luminosity collision scheme with the KLOE-2 detector introduces new challenges in terms of Interaction Region (IR) layout and optics, beam acceptance and coupling correction. The IR magnetic layout, see Fig. 2, has been designed in order to maximize the beam stay clear letting the beam trajectory pass as much as possible through the center of the magnetic elements. The IR optics provides the prescribed low-β parameters at the IP ($\beta_x$ = 0.265 m, $\beta_y$ = 0.0085 m, $\alpha_x = \alpha_y$ = 0.0, $\eta_x = \eta'_x$ = 0.0), matching at the same time the ring original layout in the arcs, as well as the phase advance between the IP and the CW sextupoles. Transverse coupling compensation is achieved by means of two anti-solenoids for each beam, and by rotating some quadrupoles in the IR.

All these aspects have been studied in details and the results are presented a discussed in a dedicated paper [10].

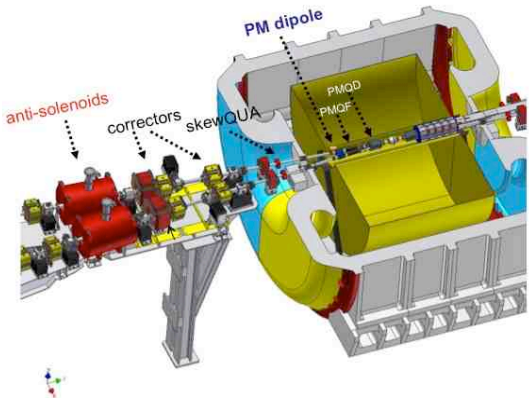

Figure 2: The KLOE-2 detector and its interaction region.

## IMPROVING BEAM DYNAMICS

The limiting factors observed during the run for the CW test required many developments related to ring impedance, lattice non-linearities, instabilities mitigation and e-cloud formation. These are meant to achieve higher currents, especially for the positrons, with better lifetime and dynamic aperture.

### Wiggler

The wiggler magnets have been modified once again to pursue the fight against non-linear terms in the magnetic field (B) started in 2001 [11].

The longitudinally and horizontally shimmed plates added on the poles in 2004 [12], which led to an improvement by a factor 2 in the dynamic aperture and in the energy acceptance [13], have been removed and a new approach has been defined in order to:

- reduce the higher order components in B,
- increase the maximum B at a given current,
- cancel the field integral on the beam trajectory.

It consists in changing the pole disposition in a such a way to keep the beam trajectory as much as possible centered with respect to the pole face [14]. A mathematical analysis shows how even terms in B, having the same symmetry as the field, still approximately cancel each other because the B sign changes from pole to pole. On the opposite odd terms, having the same symmetry as the B derivative tend to cancel inside each pole. The new configuration with shifted poles has been tested on a spare wiggler with a 37 mm pole gap, the same used for shimmed pole wigglers. A 1.726 T peak field has been measured at 550 A; reducing the current at 450 A the field becomes 1.644 T, a value still higher than the one achieved at 550 A, during the past DAΦNE operations. A further improvement has been obtained by short-circuiting one out of the five windings in the terminal poles coils; in fact the field integral of the whole wiggler can be perfectly compensated powering in series the central and terminal coils. This approach allows to get rid of eight power supplies reducing the wall plug power. The eight wigglers installed on the DAΦNE collider have been all removed, modified and measured. The residual field integral has been experimentally compensated for each device below 1 Gm by tuning the end pole clamps aperture.

### Beam stabilization

Each DAΦNE ring is equipped with three independent sophisticated bunch-by-bunch feedbacks.

In the longitudinal plane a new system is going to replace the 15 years old one. Its main characteristic consists in being affected by a lower noise level in detecting and damping the beam longitudinal oscillations. The vertical feedback will be equipped with a 12-bit ADC based hardware in order to reduce the impact of the quantization noise on the system gain. Concerning the horizontal feedback, the one installed on the positron ring deserves special attention; in fact it must keep under control a fast horizontal instability due to the e-cloud limiting the maximum positron current and, in turn, the peak luminosity. During the past run this effect has been mitigated by adding a second horizontal feedback system kicking the beam by using two out of the four injection kickers striplines powered by spare hardware. In this way ~1.2 A of positron have been stored in a stable beam with the design transverse beam dimension [15].

In view of the KLOE-2 run a more systematic approach has been undertaken to cope with the e-cloud driven instability. The transverse horizontal feedback power has been doubled (500 W now) providing ~40% increase in the kick strength which scales as the square root of the power itself. The horizontal feedback kicker has been replaced with a device with a double stripline length and reduced plate separation, providing larger shunt impedance at the low frequencies typical of the unstable modes. Moreover the kicker has been moved in a lattice position having a higher $\beta_x$ value.

However the best way to overcome the threshold in the positron current consists in avoiding the e-cloud formation; for this reason stripline electrodes have been designed and inserted in wiggler and dipole vacuum chambers of the positron ring relying on beam studies [16] showing a clear dependence of the e-cloud instability grow-rate on the beam orbit in those magnets. The e-cloud electrodes [17] consist of Cu strips, equipped with dielectric (shapal) contacts, having 2 mm total thickness, 50 mm width and 1.5 m length. An almost complete neutralization of the emitted photo-electrons [18] is expected to be achieved by applying to the electrodes a moderate dc voltage of the order of 0.5 kV.

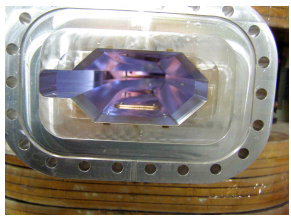

Figure 3: The e-cloud clearing electrode installed inside a dipole beam pipe, the strip shape is bent to follow the beam trajectory.

### RF FBK

A direct RF feedback system in the low level RF is being developed. This allows to reduce the cavity detuning angle, increasing the overall efficiency and limiting the reduction of the coherent '0-mode' synchrotron frequency with beam current.

### Impedance budget

The leftover old-style bellows relying on shields implemented by contiguous mini bellows have been replaced with new ones having lower impedance and providing long lasting shield contour uniformity when compressed.

The ion clearing electrodes still present in the electron ring and no longer used have been removed.

The Collimator rectangular vacuum chambers, (20 mm high and 90 mm wide), have been replaced by square ones (55 mm) to reduce their contribution to the ring impedance. In this way it becomes also possible to move the blades closer to the beam improving their effectiveness in intercepting the background otherwise hitting the experimental detector.

## OTHER HARDWARE DEVELOPMENTS

Several other systems have been improved to fulfil the requirements set by the KLOE-2 detectors and to achieve more reliable operating conditions.

The cryogenic plant has been maintained and provided with new transfer lines to cool the 4 superconducting anti-solenoids installed in the IR.

The Linac gun operating since 15 years has been replaced with a new one. A new accelerating section is going to be added at the end of the machine to make operation less critical during the positron beam injection.

The new kicker developed for the transverse horizontal positron feedback has been also used as a beam dumper. It has been installed in the opposite section with respect to the IR and will allow to dump the beam in a controlled way reducing the radiation level in the area and avoiding dangerous detector trips.

The Control System functionality has been extended to all new collider elements, and the system itself is undergoing a deep upgrade; in fact the low level control boards — based on 68040 processors and MacOS operating system — are going to be progressively replaced by Intel boards running under Linux.

## CONCLUSIONS

The developments on the DA$\Phi$NE collider have been completed in a six months shutdown. The KLOE-2 experiment is in place as well as its new IR. The detector has been already cooled down to 4.4 $^0$K. Several accelerator components and subsystems have been modified looking for higher and more stable currents having longer lifetime and for more reliable operations. The DA$\Phi$NE and KLOE-2 commissioning will start in the next few weeks.

## ACKNOWLEDGEMENTS

The activities on DA$\Phi$NE and KLOE-2 have been successfully finished on schedule thanks to the commitment of the Accelerator and Technical Divisions staff.